\documentclass{rrparticle}
\usepackage{graphicx}
\usepackage{epstopdf}
\usepackage{braket}

\newcommand{\miktex}{\hbox{Mik\kern-.15em\TeX}}

\title{Study of electron capture rates on chromium isotopes for core-collapse
simulations}
\author[] {Muhammad Majid and Jameel-Un Nabi} \affil{Faculty of Engineering Sciences,\\GIK
Institute of Engineering Sciences and Technology, Topi 23640, Khyber
Pakhtunkhwa, Pakistan\\ Email: {\em majid.phys@gmail.com,
jameel@giki.edu.pk} }


\begin{document}
\maketitle
\begin{abstract}
Electron capture rates on \emph{fp}-shell nuclei play a pivotal role
in the dynamics of stellar evolution and core collapse. These rates
play a crucial role in the gravitational collapse of the core of a
massive star activating the supernova explosion. As per simulation
results, capture rates on chromium isotopes have a major impact on
controlling the lepton-to-baryon fraction of the stellar core during
the late phases of evolution of massive stars. In this paper we
calculate the electron capture rates on isotopes of chromium with
mass range $42\leq A \leq 65$, including neutron-deficient and
neutron-rich isotopes. For the calculation of weak rates in stellar
matter, we used the pn-QRPA model with separable Gamow-Teller forces
and took deformation of nucleus into consideration. A recent study
proved this form of pn-QRPA to be the best for calculation of GT
strength distributions amongst the pn-QRPA models. The stellar weak
rates are calculated over a broad range of temperature $(0.01 \times
10^{9}-30 \times 10^{9} (K))$ and density $(10-10^{11}(g/cm^{3}))$
domain. We compare our  electron capture rates with the pioneering
calculation of Fuller, Fowler, and Newman (FFN) and with the
large-scale shell model (LSSM) calculation. Our electron capture
rates are enhanced compared to the  FFN and shell model rates.
\end{abstract}

\section{Introduction}
The Gamow-Teller (GT$_\pm$) transitions are the most useful nuclear
weak decay processes of the spin-isospin ($\sigma\tau_\pm$) type.
These transitions are not only important in nuclear physics, but
also in astrophysics where they play a decisive role in
nucleosynthesis and in supernova explosions \cite{Bet79}. The weak
interactions strongly effect the late evolution phases of massive
stars. Their operation controls the ratio of electron to baryon
($Y_e$) content of stellar matter and core entropy of the
presupernova star and thus its Chandrasekhar mass that is
proportional to Y${_e}^{2}$ \cite{Cha39}. The electron capture
decreases the number of available electrons for pressure support,
whereas $\beta$-decay proceeds in the reverse direction.
(Anti)neutrinos are produced in both processes, and escape from the
stars having densities less than 10$^{11}$ g/cm$^{3}$ thereby
carrying out entropy and energy away from the core. The
$\beta$-decay and electron capture rates are dominated by Fermi and
GT transitions. Whereas the behavior of Fermi transitions (only
significant in $\beta$-decays) is simple, an accurate description of
GT strength is a complex problem. Nuclei are completely ionized in
the stellar environment and as a result continuum electron capture
from the degenerate electron plasma occurs. The electron energies
are high enough to persuade transitions to GT resonance. The
$\beta$-decay and electron capture happen during the time of
hydrostatic burning phases. Furthermore, during final astrophysical
evolutionary phases, the importance of these processes increases
when the temperature and density of the core become huge and the
rising Fermi energy of the electrons makes the capture favorable
\cite{ffn80}.

For \emph{fp}-shell nuclides the GT transitions are considered
extremely essential for supernova physics \cite{ffn80, sad15}. GT
transitions in chromium isotopes have a special mention as per
simulation results of presupernova evolution of massive stars (e.g.
\cite{auf94, Heg01}). Not many measurements of GT strength in
chromium isotopes have been carried out to the best of our
knowledge. Zioni et al. (1972) first studied the decay of $^{46}$Cr,
by using the $^{32}$S($^{16}$O, 2n) reaction to create $^{46}$Cr
\cite{zio72}. Onishi and collaborators (2005) examined the beta
decay of $^{46}$Cr to the 1$^{+}$ states at 993 keV excitation
energy in $^{46}$V. The T=1 nucleus decays to T=0 and 1$^{+}$ levels
of daughter nucleus, termed as favored-allowed GT transitions having
small \emph{ft}-value. This experiment was done at RIKEN accelerator
research facility. Fujita et al. (2011) performed a
$^{50}$Cr($^{3}$He,t)$^{50}$Mn reaction experiment and made
measurements up to 5 MeV in daughter nuclei \cite{fuj11}. But it was
Adachi and collaborators (2007) who performed a high resolution
$^{50}$Cr($^{3}$He,t)$^{50}$Mn measurement at 0$^\circ$ and at
incident energies of 140 MeV per nucleon in order to study GT
transitions precisely. In this experiment the authors measured GT
strength up to 12 MeV in $^{50}$Mn \cite{ada07}.

There is a need to obtain more experimental data on GT strength in
\emph{fp}-shell nuclei. Next-generation radioactive ion-beam
facilities (e.g. FAIR (Germany), FRIB (USA) and FRIB (Japan)) are
expected to provide us measured GT strength distributions of many
more nuclei. It is also important to examine the GT strength in
exotic nuclide close to proton and neutron drip lines. Simulation of
stellar events require  GT strength distributions, preferably for
hundreds of nuclei. Because of scarcity of experimental data, one is
compelled to calculate GT strength distributions for \emph{fp}-shell
nuclei using a microscopic nuclear model\cite{nab13}. It is also
essential to note that in the \emph{fp}-shell nuclides, the GT
strength is distributed over several discrete states (i.e. strength
is fragmented) and in presupernova models the information of these
low-lying strengths is very significant for a precise time evolution
of $Y_e$ \cite{ffn80,auf94}. The knowledge of measured GT strength
should be broadened and theoretical attempts should be done to
reproduce them and calculate strengths of nuclei far from line of
stability \cite{nabi07,nab07}.

The first extensive attempt for the calculation of weak rates at
high densities and temperatures was  performed by Fuller, Fowler,
and Newman (FFN) \cite{ffn80}. FFN estimated these weak rates over a
wide density and temperature range i.e. $10 \leq \rho Y_e \, (\text{g cm}^{3}) \leq 10^{11}, \quad 10^{7} \leq T(\text{K}) \leq 10^{11}$
 The
excitation energies and GT strengths were calculated by employing a
zeroth-order shell model. The experimental results available at that
time were also incorporated. The FFN work for \emph{fp}-shell nuclei
was later extended by \cite{auf94}. We present here the microscopic
calculation of GT strength distributions and electron capture rates
for 24 chromium isotopes ($^{42}$Cr - $^{65}$Cr) using the
proton-neutron quasiparticle random phase approximation (pn-QRPA)
model. This model \cite{hal67, kru84, muto89} is proven to be a very
good microscopic model for the $\beta$-decay half-life estimations
far from the stability \cite{muto89, sta89}. The pn-QRPA theory was
also effectively used in the past for the calculation of
EC/$\beta^{+}$ half lives and was found to be in excellent
comparison with the experimental half-lives \cite{hir93}. The
pn-QRPA model was later modified to treat nuclear excited state
transitions \cite{muto92}. In view of successfully calculating the
terrestrial decay rates, Nabi and Klapdor-Kleingrothaus, for the
first time, used this theory to calculate the stellar weak rates and
energy losses in stellar matter for sd- \cite{nab99} and
fp/fpg-shell nuclei \cite{nab04}. Reliability of the calculated
stellar weak rates using the pn-QRPA model was discussed in detail
in Ref. \cite{nab04}. There the authors compared the pn-QRPA
calculations with measured data for thousands of nuclei and obtained
satisfactory results (see also \cite{nabi99}). A recent study took
six different pn-QRPA models to study the GT strength distributions
in chromium isotopes \cite{sad15}. The authors concluded that the
current pn-QRPA model was the best amongst all the other pn-QRPA
models and reproduced well the available experimental data and also
possessed the best predictive power for estimation of half-lives for
unknown nuclei. Another benefit of using the pn-QRPA theory is that
large configuration spaces can be handled, which are not possible in
 shell model calculations. Consequently we decided to use the same
model (along with same model parameters) to calculate stellar
electron capture rates on isotopes of chromium.

The theoretical formalism used for the calculation of electron
capture rates is briefly discussed in next section. In Sect.~3 the
calculated weak rates are presented and also compared with previous
calculations. Finally the conclusions are drawn in Sect.~4.

\section{Theoretical Formalism}
The formalism used to determine the GT strength and stellar electron
capture rates using the pn-QRPA theory is reviewed in this section.
We made the following assumptions in the calculations.

(i) Only super-allowed Fermi and GT transitions were calculated. The
contributions from forbidden transitions were presumed relatively
negligible.

(ii) The temperature was assumed high enough so that the electrons
were not bound to the nucleus anymore and obeyed the Fermi-Dirac
distribution. At kT $>$ 1 MeV, positrons come out via
electron-positron pair creation, and take the same energy
distribution function as the electrons.

(iii) The distortion of electron wave function due to the Coulomb
interaction with nucleus was represented by the Fermi function in
the phase space integrals.

(iv) Neutrinos and antineutrinos escaped freely from the stellar
interior, and their captures were not considered.

To start our calculations the Hamiltonian was chosen as
\begin{equation} \label{GrindEQ__1_}
H^{QRPA} =H^{sp} +V^{pair} +V_{GT}^{ph} +V_{GT}^{pp},
\end{equation}
where $H^{sp} $ represents the single-particle Hamiltonian,
$V^{pair} $ denotes the pairing force, $V_{GT}^{ph}$ represents the
particle-hole ($ph$) GT force and $V_{GT}^{pp}$ is the particle
particle ($pp$) GT force. Single particle energies and wave
functions were calculated in the Nilsson model \cite{nil55}, in
which the nuclear deformations was considered. Pairing among
nucleons was incorporated within the BCS approximation. The
proton-neutron residual interactions appear in two different forms,
i.e. $ph$ and $pp$ interactions, characterized by two interaction
constants $\chi$ and $\kappa$, respectively. The selections of these
two constants were done in an optimal fashion to reproduce available
experimental data and fulfilment of model independent Ikeda sum rule
\cite{isr63}. In this paper, we chose the value of $\chi$ to be
4.2/A, showing a $1/A$ dependence \cite{Hom96} and $\kappa$ equal to
0.10. Other parameters necessary for electron capture calculations
are the nuclear deformations, Nilsson potential parameters, the
pairing gaps, and the Q-values. Nilsson-potential parameters were
chosen from \cite{rag84} and the Nilsson oscillator constant was
taken as $\hbar \omega=41A^{-1/3}(MeV)$, the same for neutrons and
protons. The computed half-lives depend just weakly on the pairing
gaps values \cite{hir91}. Therefore, the conventional values of
\[\Delta _{p} =\Delta _{n} =12/\sqrt{A} (MeV)\]
were applied in this work.  Experimentally adopted values of the
deformation parameters, for even-even isotopes of chromium
($^{48,50,52,54}$Cr), extracted by relating the measured energy of
the first $2^{+}$ excited state with the quadrupole deformation,
were taken from \cite{Ram87}. For other cases the deformation of the
nucleus was calculated as
\begin{equation}
\delta = \frac{125(Q_{2})}{1.44 (Z) (A)^{2/3}},
\end{equation}
where $Z$ and $A$ are the atomic and mass numbers, respectively, and
$Q_{2}$ is the electric quadrupole moment taken from M\"{o}ller and
Nix \cite{Moe81}. Q-values were taken from the recent mass
compilation of Audi and collaborators \cite{aud12}.

The electron capture (EC) and positron decay (PD) rates from the
parent nucleus \emph{ith} state to the daughter nucleus \emph{jth}
state is specified by
\begin{equation}
 \lambda _{ij}^{EC(PD)} =\ln 2\frac{f_{ij}^{EC(PD)} (T,\rho ,E_{f})}{(ft)_{ij} }
\end{equation}
where $(ft)_{ij}$ is connected to the reduced transition probability
($B_{ij}$) by
\begin{equation}
(ft)_{ij} =D/B_{ij}
\end{equation}
where D is constant and is given as
\begin{equation}
 D=\frac{2\ln 2\hbar ^{7}\pi^{3}}{g_{v}^{2} m_{e}^{5}
c^{4} }
\end{equation}
and $B_{ij}$ is given by
\begin{equation}
 B_{ij} = B(F)_{ij} +((g_{A}/g_{V})^{2} B(GT)_{ij}
\end{equation}
We took the value of D = 6295 s \cite{hir93} and $g_{A}/g_{V}$ as
-1.254. The reduced transition probabilities B(F) and B(GT) are
specified by
\begin{equation}
B(F)_{ij}=\frac{1}{2J_{i} +1} \langle{j}\parallel\sum\limits_{k}
t_{\pm}^{k}\parallel {i}\rangle|^{2}
\end{equation}
\begin{equation}
B(GT)_{ij}=\frac{1}{2J_{i} +1} \langle{j}\parallel\sum\limits_{k}
t_{\pm}^{k}\overrightarrow{\sigma}^{k}\parallel {i}\rangle|^{2}
\end{equation}
here $\overrightarrow{\sigma}(k)$ is the spin operator and $t_{\pm
}^{k}$ represent the isospin raising and lowering operator. For
construction of parent and daughter excited states and calculation
of nuclear matrix elements we refer to \cite{nab99}. The phase space
integral $f_{ij}$ is an integral over total energy. For electron
capture it is given by
\begin{equation}
f_{ij}^{EC} = \int _{w_{l} }^{\infty }w\sqrt{w^{2} -1}  (w_{m}
+w)^{2} F(+ Z,w)G_{- } dw,
\end{equation}
and for positron emission
\begin{equation}
f_{ij}^{PD} = \int _{1} ^{w_{m} }w\sqrt{w^{2} -1}  (w_{m} -w)^{2}
F(- Z,w)(1- G_{+ }) dw.
\end{equation}
Here $w$ represents the total energy of the electron including its
rest mass, and $w_{l}$ denotes the total capture threshold energy
(rest + kinetic) for electron capture. $G_{-} (G_{+})$ is the
electron (positron) distribution functions.
\begin{equation}
 G_{-} =\left[\exp \left(\frac{E-E_{f} }{kT} \right)+1\right]^{-1}
\end{equation}
\begin{equation}
G_{+} =\left[\exp \left(\frac{E+2+E_{f} }{kT} \right)+1\right]^{-1}
\end{equation}
Here $E=(w - 1)$ is the kinetic energy of the electrons, $E_{f}$ is
the Fermi energy of the electrons, T is the temperature, and $k$ is
the Boltzmann constant. The Fermi functions $F(Z,w)$  are determined
according to the procedure used by \cite{gov71}. If the
corresponding electron or positron emission total energy $(w_{m})$
is greater than -1, then $w_{l} = 1$, and if less than or equal to
1, then $w_{l}=|w_{m}|$, where $w_{m}$ is the total $\beta$ decay
energy,
\begin{equation} w_{m} = m_{p} -m_{d}+E_{i} -E_{j}
\end{equation}
\noindent where $m_{p}$ ($m_{d}$) and $E_{i}$ ($E_{j}$) are mass and
excitation energies of the parent (daughter) nucleus respectively.
The number density of electrons linked with protons and nuclei is
$\rho Y_{e}N_{A}$ ( where $\rho$ is the baryon density and $N_{A}$
is Avogadro number)
\begin{equation}
\rho Y_{e}=\frac{1}{\pi ^{2} N_{A} }(\frac{m_{e}c}{\hbar})^{3}\int
_{0}^{\infty }(G_{-}  -G_{+} )p^{2} dp
\end{equation}
where $p = (w^{2}-1)^{1/2}$ is the momentum of electron. Eq. 14 was
used for an iterative calculation of Fermi energies for selected
values of T and $Y_{e}$. As the temperature in the interior of stars
is very high so there is a finite probability of occupation of
parent excited states. The  total EC and PD rates per unit time per
nucleus is
\begin{equation}
\lambda^{EC(PD)} =\sum _{ij}P_{i} \lambda _{ij}^{EC(PD)},
\end{equation}
where $P_{i}$ follows the normal Boltzmann distribution. In Eq. 15,
the summation was taken over all the initial and final states until
reasonable convergence was achieved in our calculated rates. The
Fermi strength is concentrated in a very narrow resonance centered
around the isobaric analogue state (IAS) for the ground and excited
states. In case of Fermi strength the isobaric analogue state (IAS)
was calculated  by operating on the associated parent states with
the isospin raising or lowering operator

 \[T_{\pm } =\sum _{i}t_{\pm } (i) ,\]
where the sum runs over the nucleons. The Fermi matrix element
depends only on the nuclear isospin (T) and its projection $T_{z}$
(equal to (Z-N)/2) for the parent and daughter nucleus. The energy
of the IAS was calculated according to \cite{gro90} and the reduced
transition probability was calculated using

\[B(F)=T(T+1)-T_{zi} T_{zf} ,\]
where $T_{zi}$  and $T_{zf}$  are the third components of the
isospin of initial and final analogue states, correspondingly.

\section{Results and Discussion}

Earlier in Ref. \cite{sad15} three different QRPA models, namely the
Pyatov method (PM), the Schematic model (SM) and the pn-QRPA model,
were considered for the calculation of GT transitions in chromium
isotopes. The idea was to study the GT transitions in different QRPA
models. These models mainly studied the effect of deformation,
particle-hole and particle-particle interactions in QRPA
calculations. It was observed that the lowest total GT strength
values were calculated by the PM. Further the PM model failed to
yield the desired fragmented GT strength distribution. The reason
was that the PM considered only spherical nuclei, due to which most
of the GT strength was concentrated in one specific state. The
biggest values of total GT strength was calculated by the SM, but it
placed the GT centroid at much higher excitation energy in daughter
nucleus. The PM also resulted in higher placement of centroid values
in the daughter nuclei. On the other hand, the pn-QRPA model placed
the centroids at lower energies in daughter which translated into
stronger weak rates in astrophysical environment.  Further it was
shown that the pn-QRPA model compared well with measured GT strength
distributions wherever available  and emerged as the best model for
calculation of GT strength distributions \cite{sad15}. In this
manuscript we use the same pn-QRPA model for the calculation of
$\beta$-decay half-lives and stellar electron capture (EC) rates. We
consider $24$ isotopes of chromium, in mass range $^{42-65}$Cr, for
the calculation of EC  rates. These nuclei include both stable
($^{50}$Cr, $^{52-54}$Cr) and unstable isotopes of chromium,
including neutron deficient and neutron rich cases. We quenched our
pn-QRPA results by a  factor of $f_{q}^{2}$ = (0.6)$^{2}$
\cite{Vet89, Gaa83} in calculation of EC rates (akin to other
microscopic calculations including shell model calculations).
Interestingly \cite{Vet89} and \cite{Roe93} predicted the same
quenching factor of 0.6 for the RPA calculation in the case of
$^{54}$Fe when comparing their measured strengths to RPA
calculations.

Fig. 1 shows that our calculated $\beta$-decay half-lives for
isotopes of chromium agree quite well with the measured values. The
experimental half-lives were taken from \cite{aud12}. As mentioned
earlier, $^{50}$Cr, $^{52}$Cr, $^{53}$Cr and $^{54}$Cr are stable
isotopes of chromium.

The total GT$_{+}$ and GT$_{-}$ strengths (represented as
$\sum$B(GT$_{+}$) and $\sum$B(GT$_{-})$ respectively, in this work)
are related to the re-normalized Ikeda sum rule (ISR$_{re-norm}$) as
\begin{equation}
ISR_{re-norm} = \sum B(GT_{-}) - \sum B(GT_{+})\cong
3f_{q}^{2}(N-Z). \label{Eqt. ISR}
\end{equation} where Z and N
represent the numbers of protons and neutrons, respectively \cite
{isr63}. Fig.~2 shows the comparison of calculated $ISR_{re-norm}$
with the model-independent theoretical predictions. It is clear from
Fig.~2 that the $ISR_{re-norm}$ is satisfied well by the pn-QRPA
model (deviates at the maximum  by only a few percent). The SM and
PM models satisfied the Ikeda Sum Rule (ISR) for even-even Cr
isotopes but showed some deviations for odd-A cases. The pn-QRPA
model fulfilled the ISR for both even-even and odd-A Cr isotopes.

The pn-QRPA calculated EC rates are shown in Table 1.  We present
these EC rates for temperatures (1, 3, 10 and 30)$\times 10^{9}K$
and at selected densities ($10^{3}$, $10^{7}$ and
$10^{11}gcm^{-3}$). It is to be noted that the calculated EC rates
(Eq. 15) are tabulated in log to base 10 values (in units of
s$^{-1}$). The EC rates increases with increasing temperature and
density (Eq. 9). The probability of occupation of parent excited
states increases with increasing stellar temperature and hence
contribute effectively to the total rates at high temperatures. It
is to be noted that the Fermi energy of electrons increases with
increasing stellar densities. This leads to substantial increment of
EC rates at high density. However at high densities the rate of
change of EC decreases with temperature. The complete set of these
rates for all isotopes of chromium can be requested from the authors
on demand.

The calculated EC rates on selected chromium isotopes $^{50}$Cr,
$^{51}$Cr, $^{53}$Cr, $^{56}$Cr and $^{57}$Cr of astrophysical
importance are shown in Fig.~3. Isotopes of chromium, namely
$^{51,53,56,57}$Cr, were included in the Aufderheide's list of key
nuclei for electron capture rates \cite{auf94}. In addition,
$^{50}$Cr, $^{51}$Cr, and $^{53}$Cr were considered  amongst the
most important nuclei for modeling of presupernova evolution of
massive stars that decrease $Y_e$ of stellar matter (for detail see
\cite{Heg01}). Graphs in Fig.~3 illustrate that the EC rates remain,
more or less, constant in low density regions. In these regions the
beta-decay compete well with capture rates before core collapse. As
the stellar core stiffens to high values $(10^{8} gcm^{-3}-10^{11}
gcm^{-3})$, the electron Fermi energy also increases thereby
increasing the EC rates.  At later phases of the collapse,
$\beta$-decay becomes insignificant as an increased electron
chemical potential, which grows like $\rho^{1/3} $ during in fall,
considerably, decreases the phase space. These high EC rates during
the collapse make the stellar composition more neutron-rich. Thus in
the final state of the collapse phase the $\beta$-decay is
relatively trivial due to Pauli-blocking of the electron phase space
\cite{nabi05}.

Large scale shell model (LSSM) was used to calculate the EC rates on
$^{45-58}$Cr \cite{lan01}. Fuller, Fowler, and Newman (FFN) \cite
{ffn80}, on the other hand, used their model to calculate EC rates
on $^{45-60}$Cr isotopes. The FFN rates had been used in many
simulation codes (e.g., KEPLER stellar evolution code) while LSSM
rates were employed in recent simulation of presupernova evolution
of massive stars in the mass range 11-40\;M$_\odot$ \cite{Heg01}.
The comparison of our results with the FFN \cite{ffn80} and large
scale shell model (LSSM) rates \cite{lan01} are shown in Figures 4
and 5. In these figures, for each isotope, we depict three panels.
In each case the upper panel shows comparison of calculated EC rates
at temperature $1\times10^{9}K$, whereas the middle and lower panels
show comparison at temperatures $10\times10^{9}K$ and
$30\times10^{9}K$, respectively. The selected values of densities
are 10$^{3}$ gcm$^{-3}$, 10$^{7}$gcm$^{-3}$ and 10$^{11}$gcm$^{-3}$
(corresponding to low, medium and high densities). Comparison of the
capture rates with previous calculations can be divided into two
categories. In first category our rates are enhanced at all
temperatures and densities compared to previous calculations by as
much as two orders of magnitude. The results are shown in Fig.~4.
Unmeasured matrix elements for allowed transitions were assigned an
average value of $log ft = $5 in FFN calculation. On the other hand
these transitions were calculated in a microscopic fashion using the
pn-QRPA theory (and LSSM) and depict a more realistic picture of the
events taking place in stellar environment. The total strengths,
centroids and widths of our calculated GT distribution can be seen
in Table~2. Both EC and positron-decay rates  are very sensitive to
the position of the GT$_{+}$ centroid. The (n,p) experiment on a
nuclide (Z, A) demonstrates the position where in (Z-1, A) the
GT$_{+}$ centroid analogous to the ground state of (Z, A) resides.
The electron capture  and $\beta^{+}$-decay are exponentially
sensitive to the placement of GT$_{+}$ resonance whereas the total
B(GT$_{+}$) affect the astrophysical rates in a more or less linear
fashion \cite{auf96}.  The widths of calculated GT distribution
provide a measure of how much the individual GT states are dispersed
around the centroid value. The total B(GT$_{+}$) strength decreases
monotonically as the mass number increases. For both even and odd
mass nuclei, the GT resonance energy for FFN cluster around 2, 4,
and 6 MeV.  The LSSM calculated centroid energies are dispersed as
the residual interaction fragment the GT strength (see Fig.~6 of
\cite{lan00}). Table~2 clearly shows that the pn-QRPA calculated
centroid energies of GT strength are also scattered due to
fragmentation of the GT strength. Compared to the LSSM centroids,
FFN place the GT resonance energy usually at higher excitation
energies for even-even nuclei and often at too low excitation
energies for odd-odd nuclei. FFN place the GT resonance energy at
around 6 MeV for odd-A nuclei having odd number of neutrons.
Compared to the pn-QRPA and LSSM calculated GT centroids, the FFN
estimate for these nuclei are too high. The EC rates calculated by
the pn-QRPA model are bigger than those calculated by FFN and LSSM
due to lower placement of centroids in our model. The pn-QRPA
calculated centroid values of B(GT$_{+}$) strength distributions for
some important nickel isotopes are shown in Table~3. For comparison
the centroids calculated by LSSM, FFN and those by Pruet and Fuller
\cite{Pru03} are also given.  In the last column  centroids  of
measured B(GT$_{+}$) strength distributions are shown. Measured data
was taken from \cite{wil95,elk94}. The pn-QRPA calculated centroids
values are in reasonable agreement with the experimental data,
except for $^{62}$Ni.

In the other category at lower stellar temperatures our rates are in
reasonable agreement with LSSM results. However in case of $^{56,
58}$Cr isotopes at low density and temperature region, our rates and
LSSM rates are enhanced by around seven orders of magnitude as
compared to FFN EC rates. At high density and low temperature region
the mutual comparison between all calculations is decent. Fig.~5
shows the mutual comparison of EC rates for this category. At high
temperatures and density region the situation is more interested if
we compare the results of LSSM and FFN. At high temperature and
density region, the LSSM rates are too small as compared to FFN and
pn-QRPA rates. The Lanczos-based approach used by LSSM and pointed
by \cite{Pru03} provides the reason for this discrepancy. The
calculated LSSM decay rates is a function of the number of Lanczos
iterations essential for convergence and this behavior of partition
functions can affect their estimates of high temperature weak rates.
Accordingly at high temperatures the LSSM rates tend to be too low.
The pn-QRPA calculation do not suffer from this convergence problem
as it is not Lanczos-based. At high temperatures, where the
occupation probability of excited states is larger, our EC rates for
all isotopes are enhanced by an order of magnitude as compared to
LSSM rates. There are several other reasons for the enhancement of
our rates. The pn-QRPA model provides adequate model space which
effectively handle all the excited states in parent as well as in
daughter nuclei.  We also do not consider the Brink's hypothesis in
our calculations to approximate the contribution from parent excited
levels. This approximation was used both by FFN and LSSM. Brink's
hypothesis states that GT strength distribution on excited states is
\textit{identical} to that from ground state, shifted \textit{only}
by the excitation energy of the state.  We carried out a
state-by-state calculation of these capture rates from parent to
daughter states in a microscopic way and added them at the end to
obtain the total EC rates (Eq.~15). It is further to be noted that
both LSSM and pn-QRPA model perform a microscopic calculation of all
energy eigenvalues and GT matrix elements for ground state of parent
nucleus. Accordingly, whenever ground state rates command the total
rate, the two calculations are found to be in excellent agreement.
For cases where excited state partial rates influence the total
rate, differences are seen between the two calculations.\\

One important question could be to know how the electron capture
rates compete with the positron decay (PD) rates for these isotopes
of chromium during presupernova evolution of massive stars. Table~4
displays the ratio of calculated EC to PD rates at selected
temperatures (1, 5, 10 and 30)$\times 10^{9}K$ and densities
($10^{7}$, $10^{9}$ and $10^{11}gcm^{-3}$). It is observed that in
$^{42-47}$Cr nuclides at stellar temperatures (1, 5, and 10)$\times
10^{9}K$  and density $10^{7}$ gcm$^{-3}$, $\beta^{+}$-decay rates
 are greater than the EC rates by 1-2 orders
of magnitude and must be taken into account in simulation codes. At
high densities ($10^{9}- 10^{11}$) g/cm$^{3}$ the EC rates are
bigger than the competing $\beta^{+}$ rates by 1-4 orders of
magnitude. As mentioned before the electron Fermi energy increases
at high densities which in turn lead to significant enhancement in
calculated electron capture rates. As $N \geq Z$, it is clear from
Table~4 that the EC rates exceed the competing $\beta^{+}$ rates
both in low and high temperature and density region. The PD values
decrease as the neutron number (N) increases. As N $\geq 31$,
calculated PD rates become less than $10^{-100}$ and are not shown
in Table~4. For all these isotopes the PD rates can safely be
neglected in comparison with the EC rates.

\section{Conclusions}
Chromium isotopes are advocated to play a vital role among the
iron-regime nuclide controlling the dynamics of core-collapse of
massive stars. The EC rates on Cr isotopes may be used as a nuclear
physics input parameter for core-collapse simulation codes.  Here we
present the calculation of stellar EC rates on twenty-four isotopes
of chromium. We considered a total of 100 parent and daughter
excited states (covered energy range was in excess of 10 MeV) for
the microscopic calculation of these rates. Our model calculation
reproduced well the measured $\beta$-decay half-life values and also
fulfilled the Ikeda Sum Rule. Later we performed calculation of
stellar electron capture and positron-decay rates of chromium
isotopes. We compared our results with the previous calculations of
FFN and LSSM. Our calculated EC rates are enhanced in the
presupernova era as compared to previous calculations and this is an
interesting finding. From astral viewpoint these enhanced EC rates
may have substantial impact on the  late stage evolution of massive
stars and the shock waves energetics. Results of simulations
illustrate that EC rates have a solid effect on the core collapse
trajectory and on the properties of the core at bounce. We urge
collapse simulators to test run our calculated EC rates in their
codes to search for possible interesting outcomes.  $\beta^{+}$
decay rates are only important for $N \leq Z$  chromium isotopes up
to stellar density of $10^{7}gcm^{-3}$. For remaining chromium
isotopes and higher temperature-density regions, the $\beta^{+}$
decay rates can safely be neglected when compared with the
corresponding EC rates.

\begin{acknowledgement}
J.-U. Nabi would like to acknowledge the support of the Higher
Education Commission Pakistan through the HEC Project No. 20-3099.
\end{acknowledgement}


\begin{figure}[h]
\includegraphics[scale=0.45]{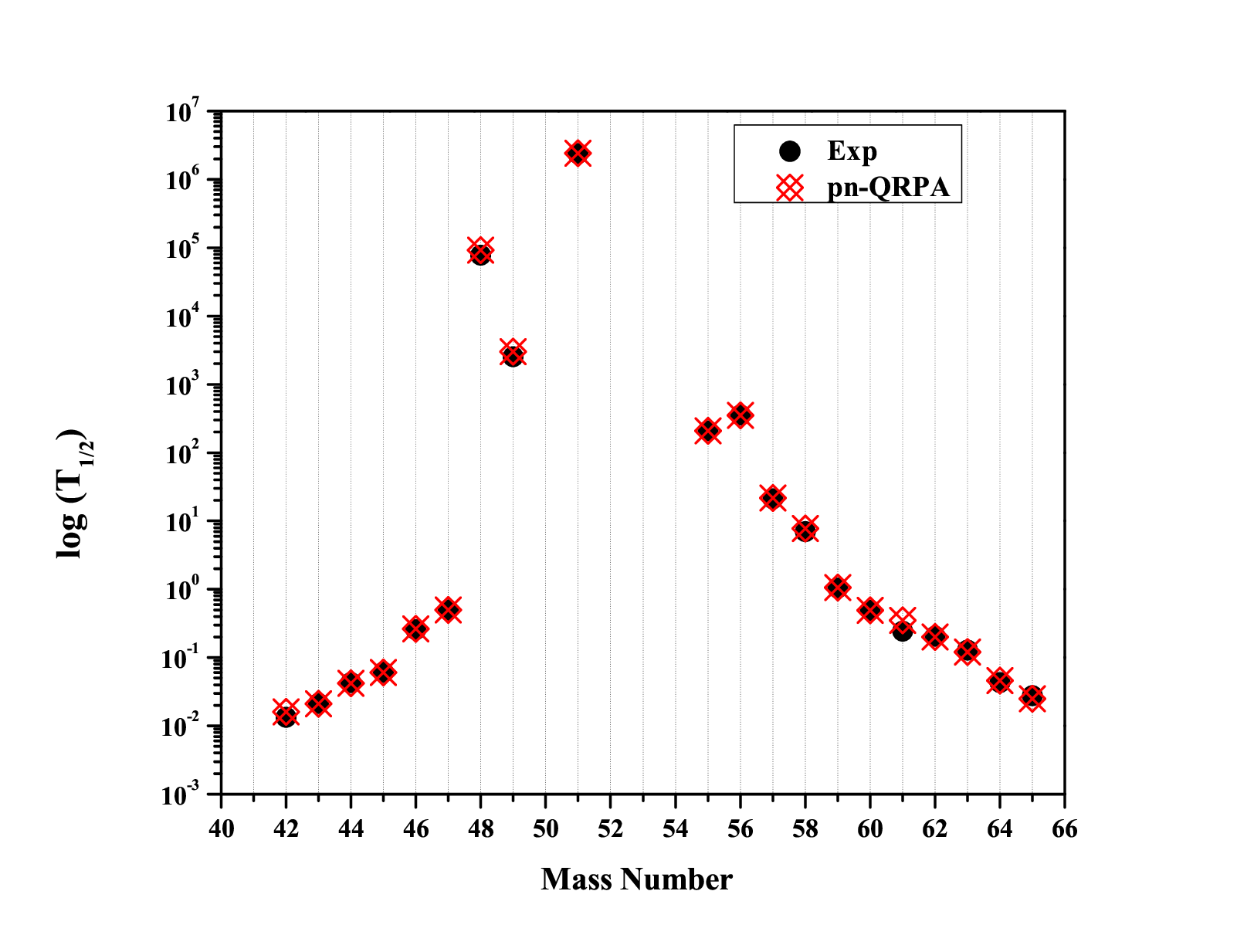}
\caption{Total $\beta$-decay half-lives for Cr isotopes calculated
from the pn-QRPA model (this work) in comparison with the
experimental data \cite{aud12}. $^{50,52-54}$Cr are stable. }
\label{fig1}
\end{figure}

\begin{figure}[h]
\includegraphics[scale=0.5]{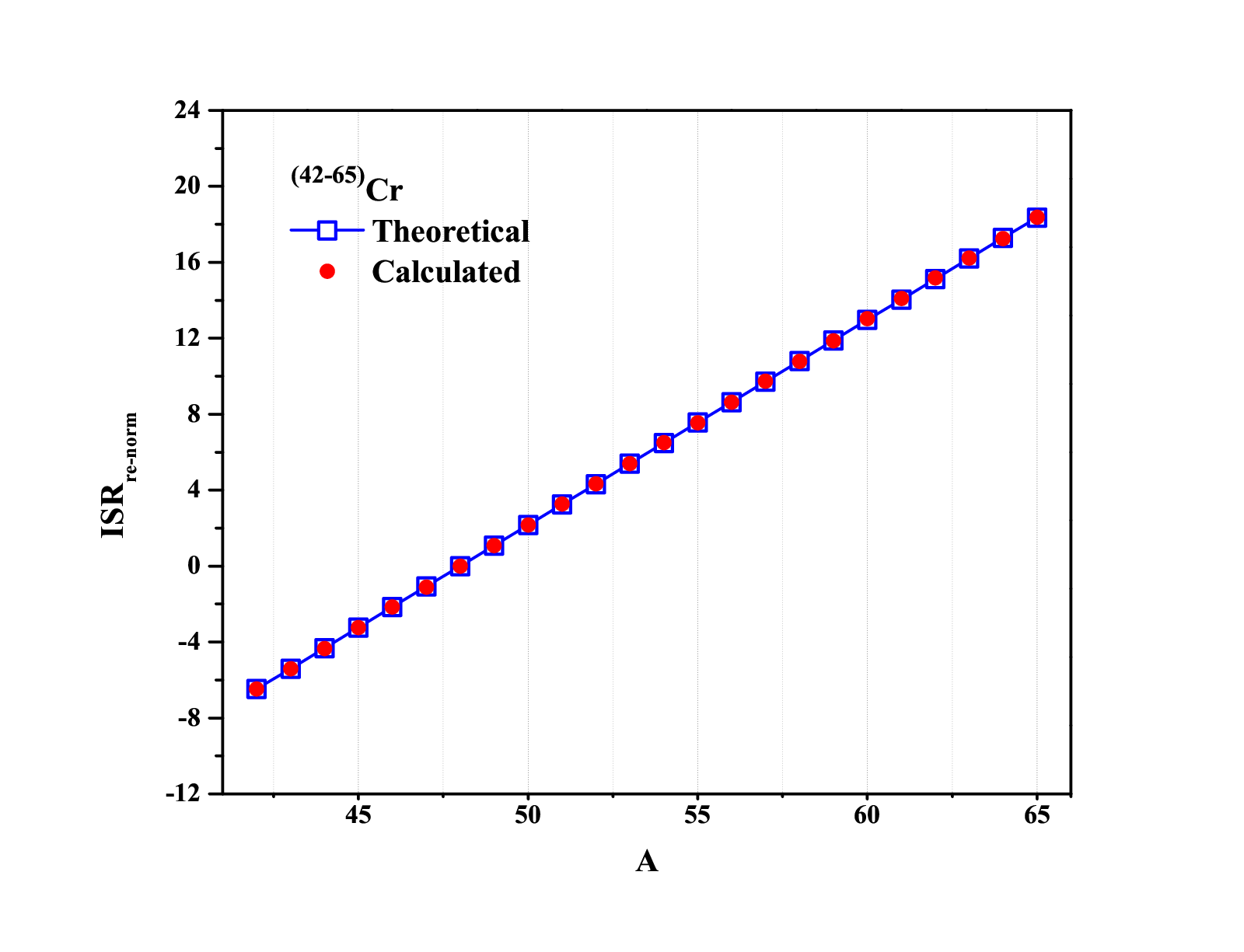}
\caption{Comparison of calculated and theoretical re-normalized
Ikeda Sum Rule.} \label{fig2}
\end{figure}

\begin{figure}[h]
\begin{center}
  \begin{tabular}{cc}
    \includegraphics[scale=0.25]{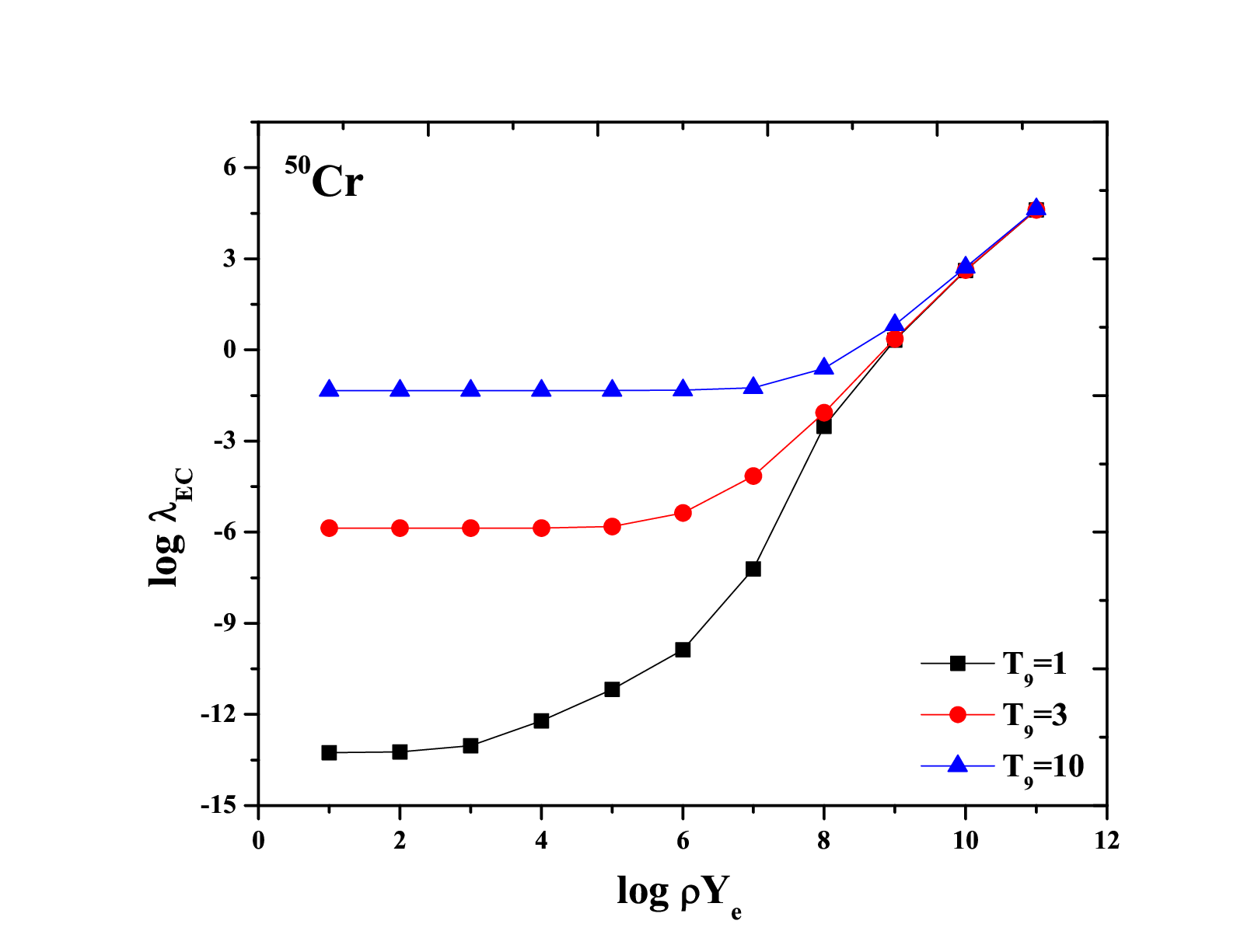} &
    \includegraphics[scale=0.25]{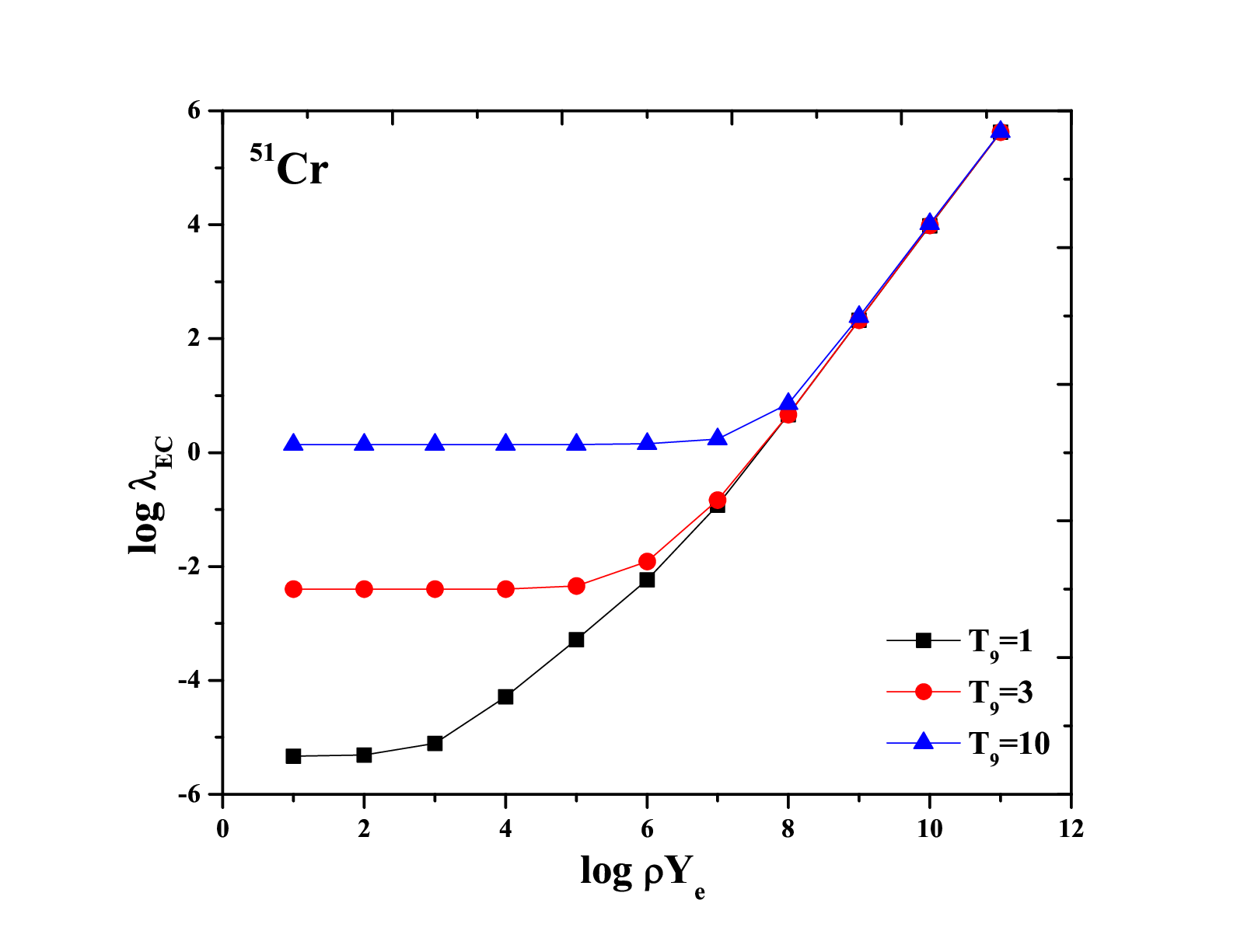}\\
    \includegraphics[scale=0.25]{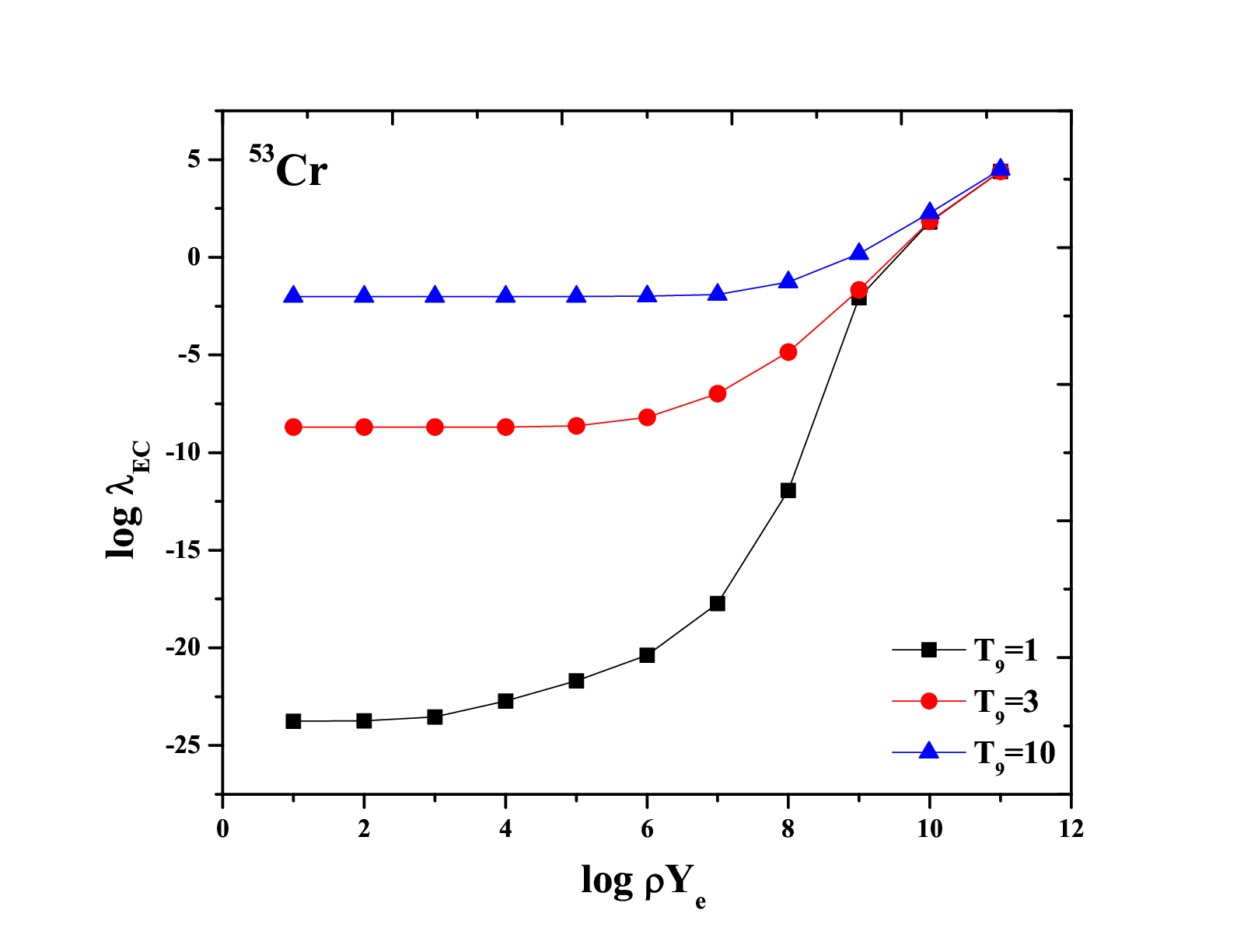} &
    \includegraphics[scale=0.25]{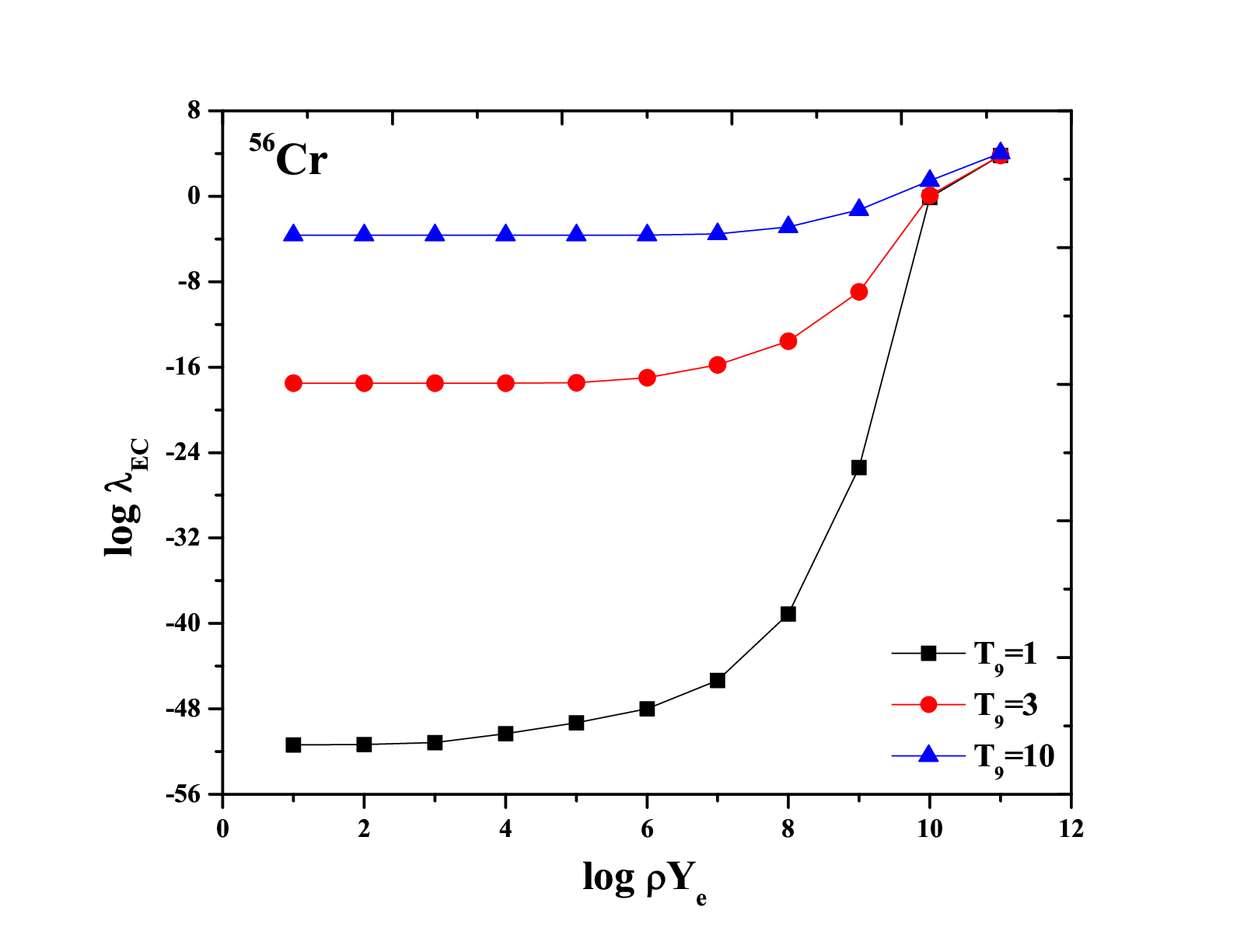}\\
\end{tabular}
    \includegraphics[scale=0.28]{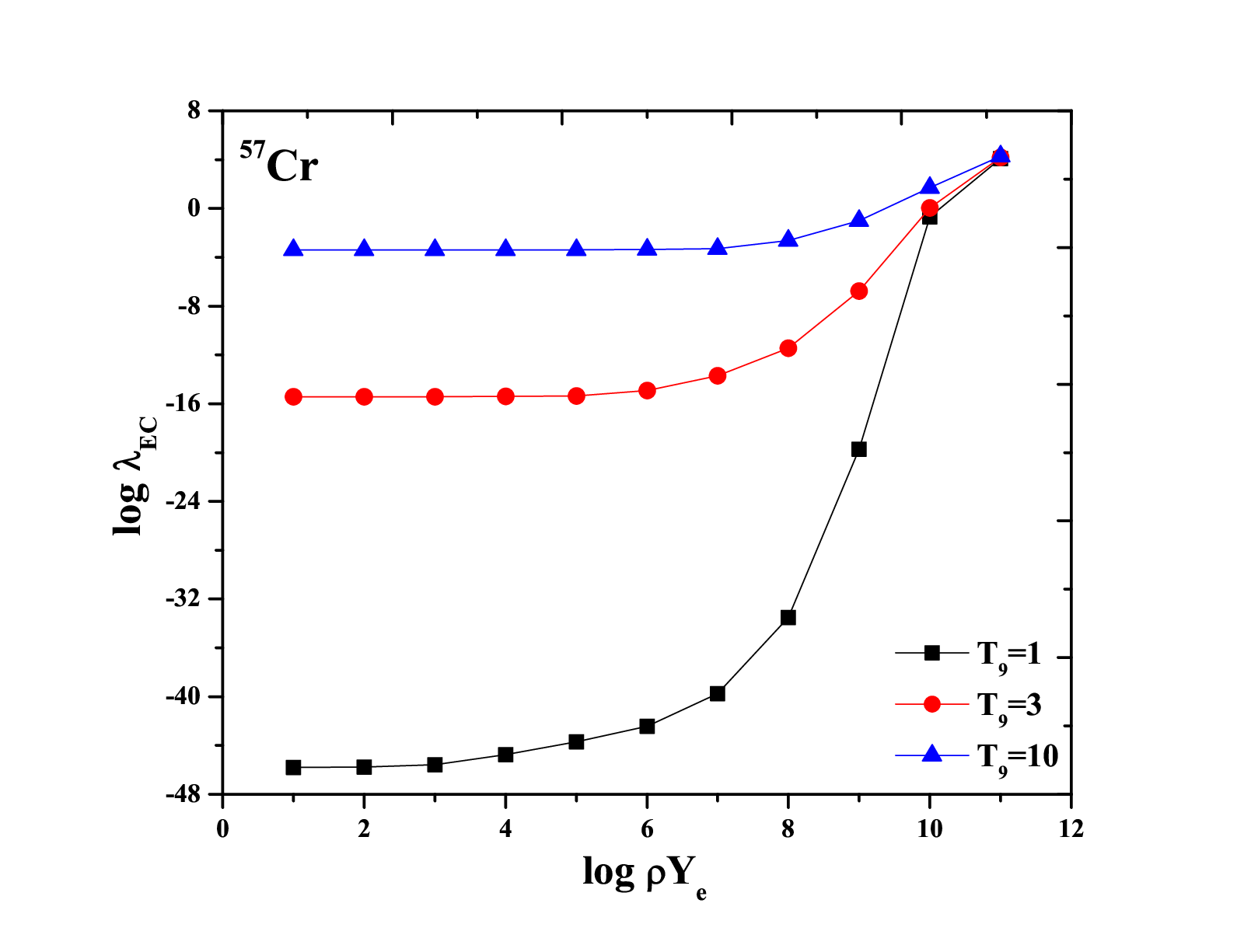}\\

\caption{Electron capture rates on ($^{50,51,53,56,57}$Cr) isotopes
as function of stellar densities ($\rho$Y$_{e}$) having units of
g/cm$^{3}$ at different selected temperatures. Temperatures
(T$_{9}$) are given in units of 10$^{9}$ K and $\log\lambda_{EC}$
represents the log (to base 10) of EC rates in units of
s$^{-1}$.}\label{fig3}
\end{center}
\end{figure}
\begin{figure}[]
\begin{center}
  \begin{tabular}{cc}
    \includegraphics[scale=0.32]{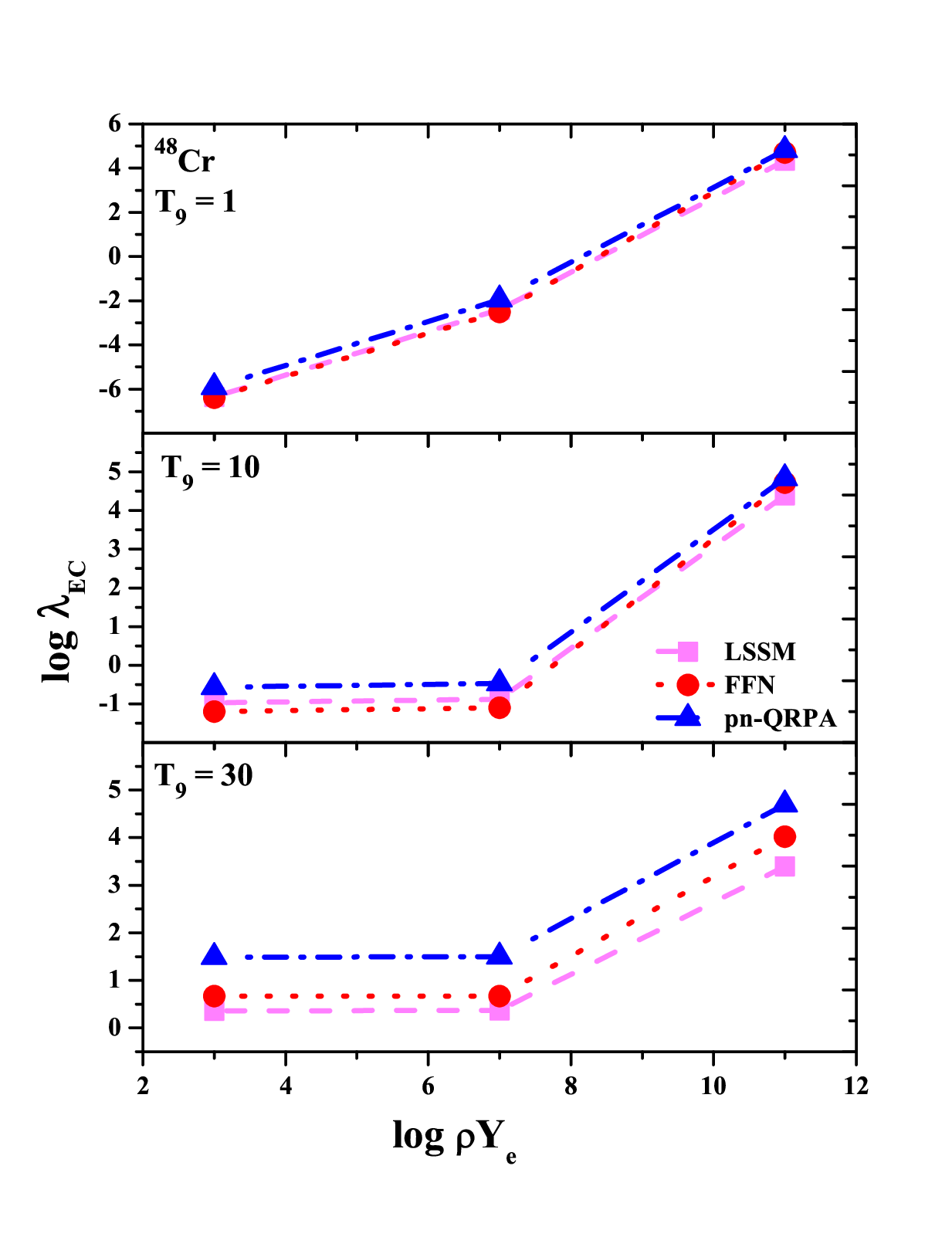} &
    \includegraphics[scale=0.32]{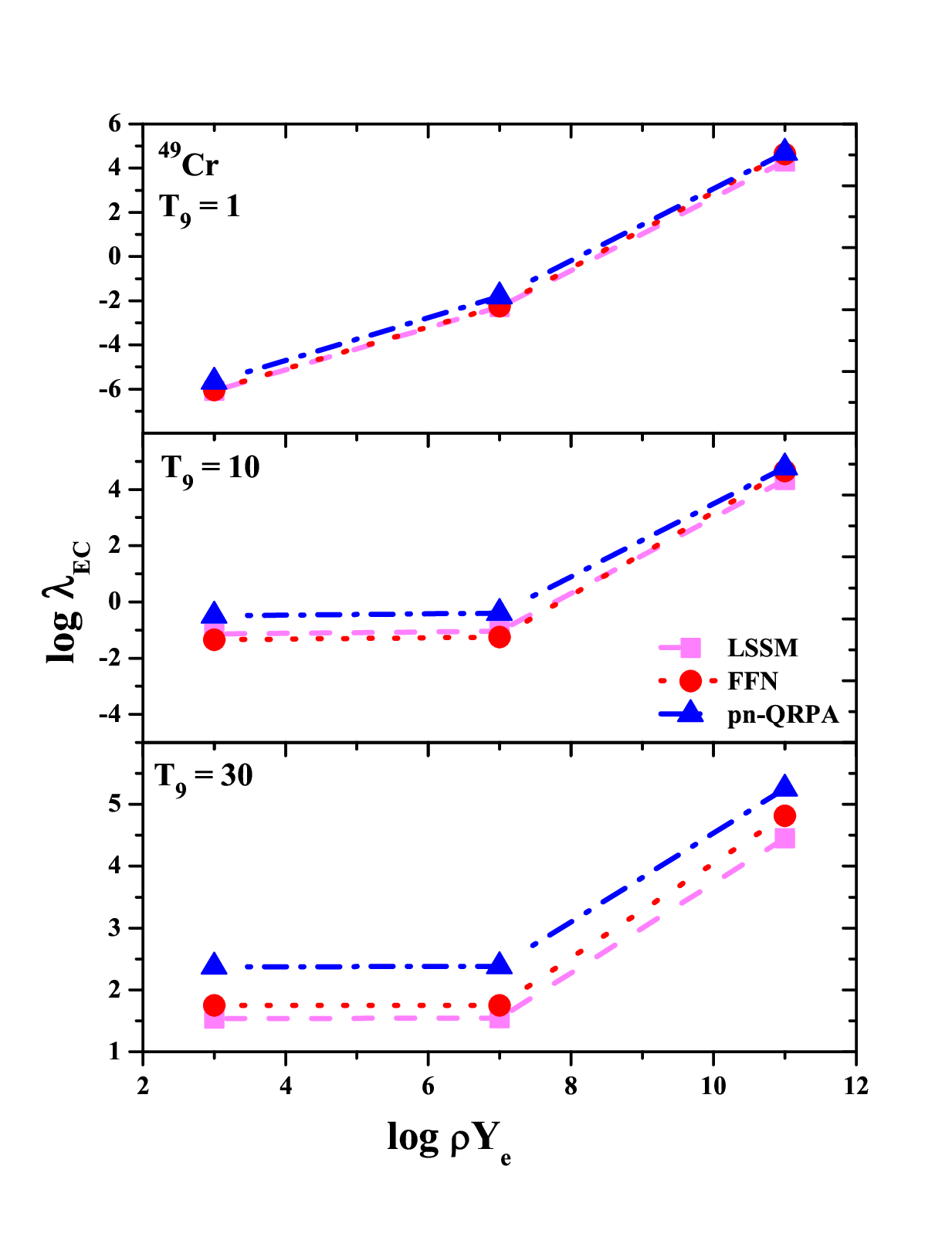}\\
    \includegraphics[scale=0.31]{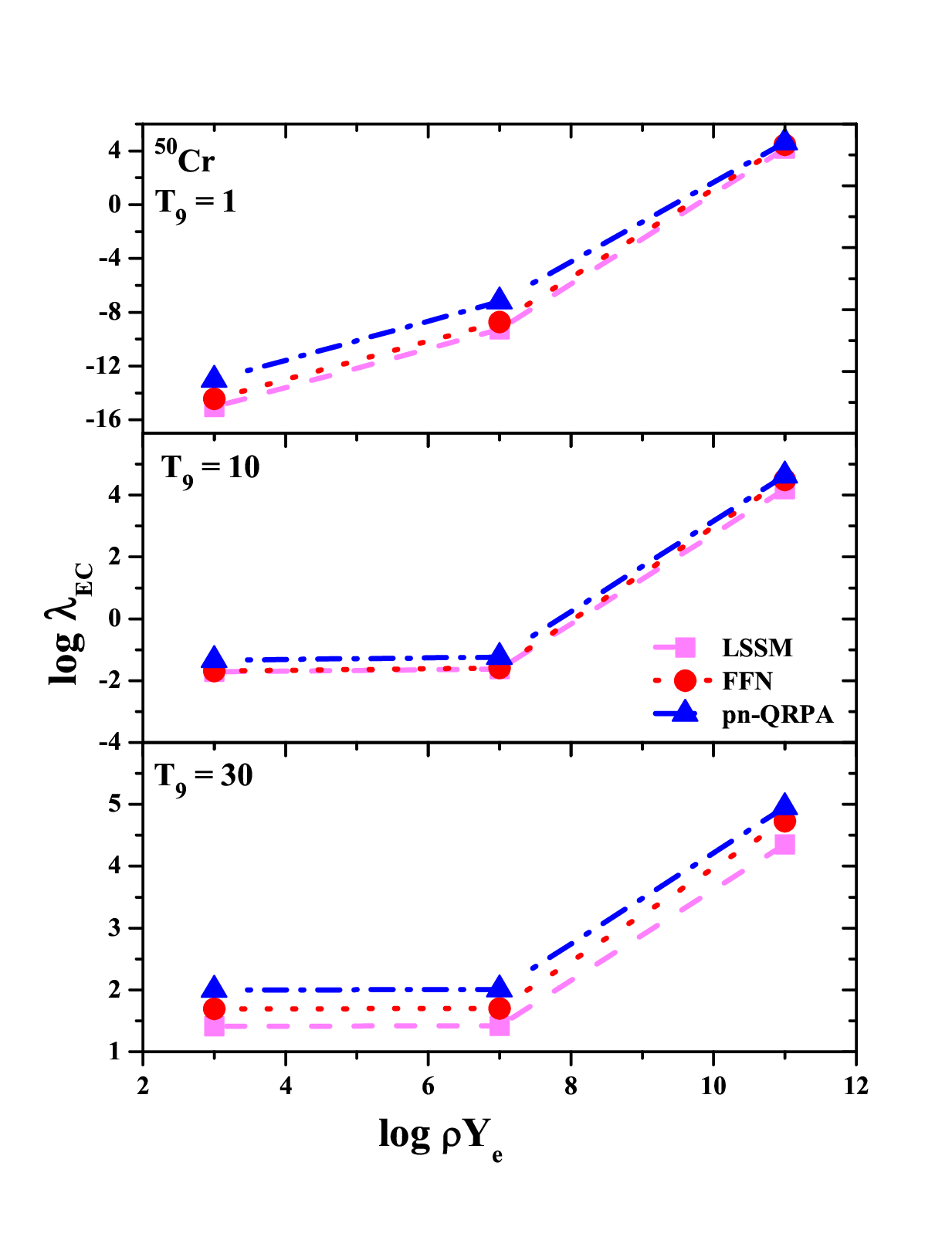} &
    \includegraphics[scale=0.31]{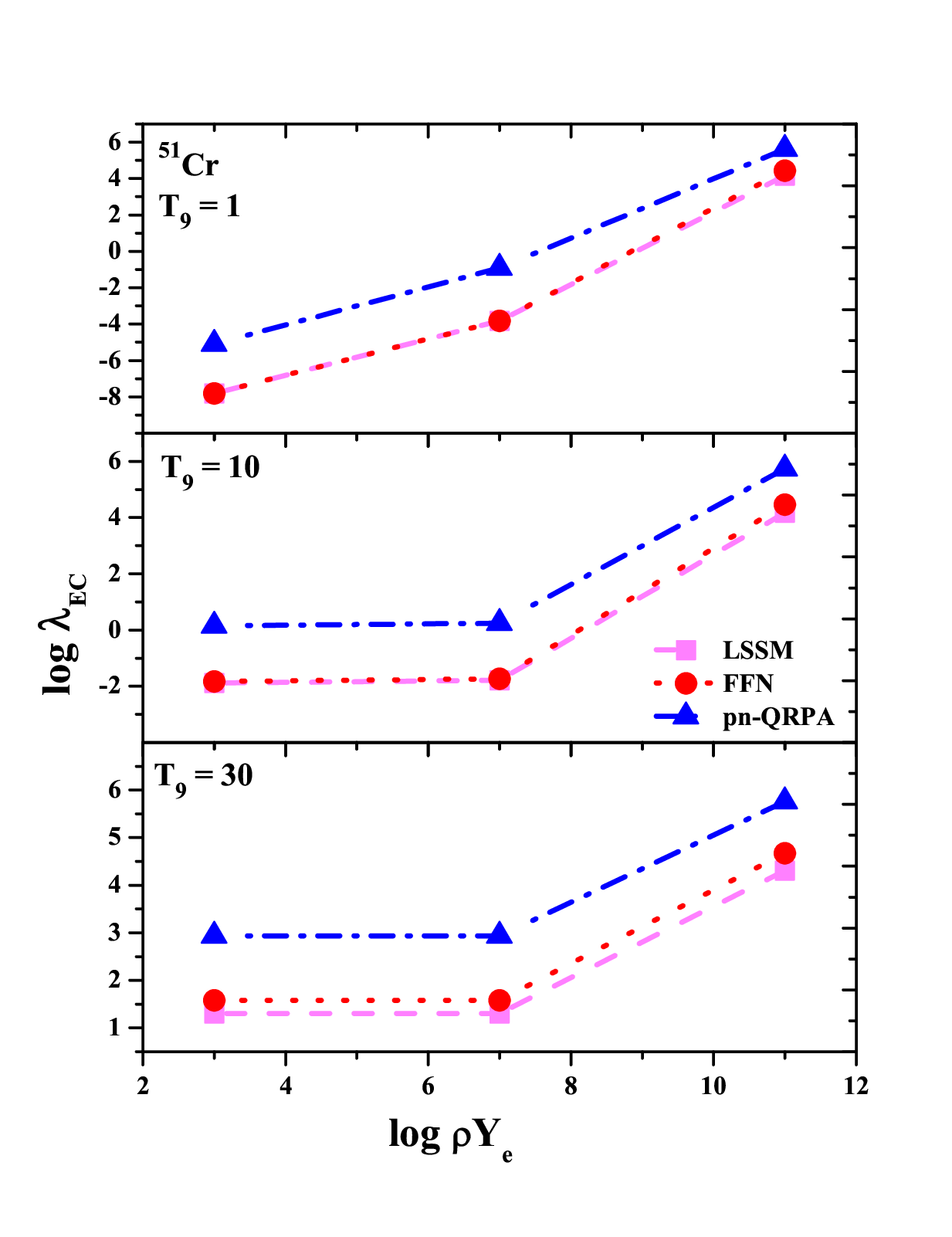}\\
\end{tabular}
\caption{Comparison of EC rates of pn-QRPA (this work) with those of
FFN \cite{ffn80} and large scale shell model (LSSM) \cite{lan01} as
function of stellar densities ($\rho$Y$_{e}$) having units of
g/cm$^{3}$ at different selected temperatures. Temperatures
(T$_{9}$) are given in units of 10$^{9}$ K and $\log\lambda_{EC}$
represents the log (to base 10) of EC rates in units of
s$^{-1}$.}\label{fig4}
\end{center}
\end{figure}
\begin{figure}[]
\begin{center}
  \begin{tabular}{cc}
    \includegraphics[scale=0.33]{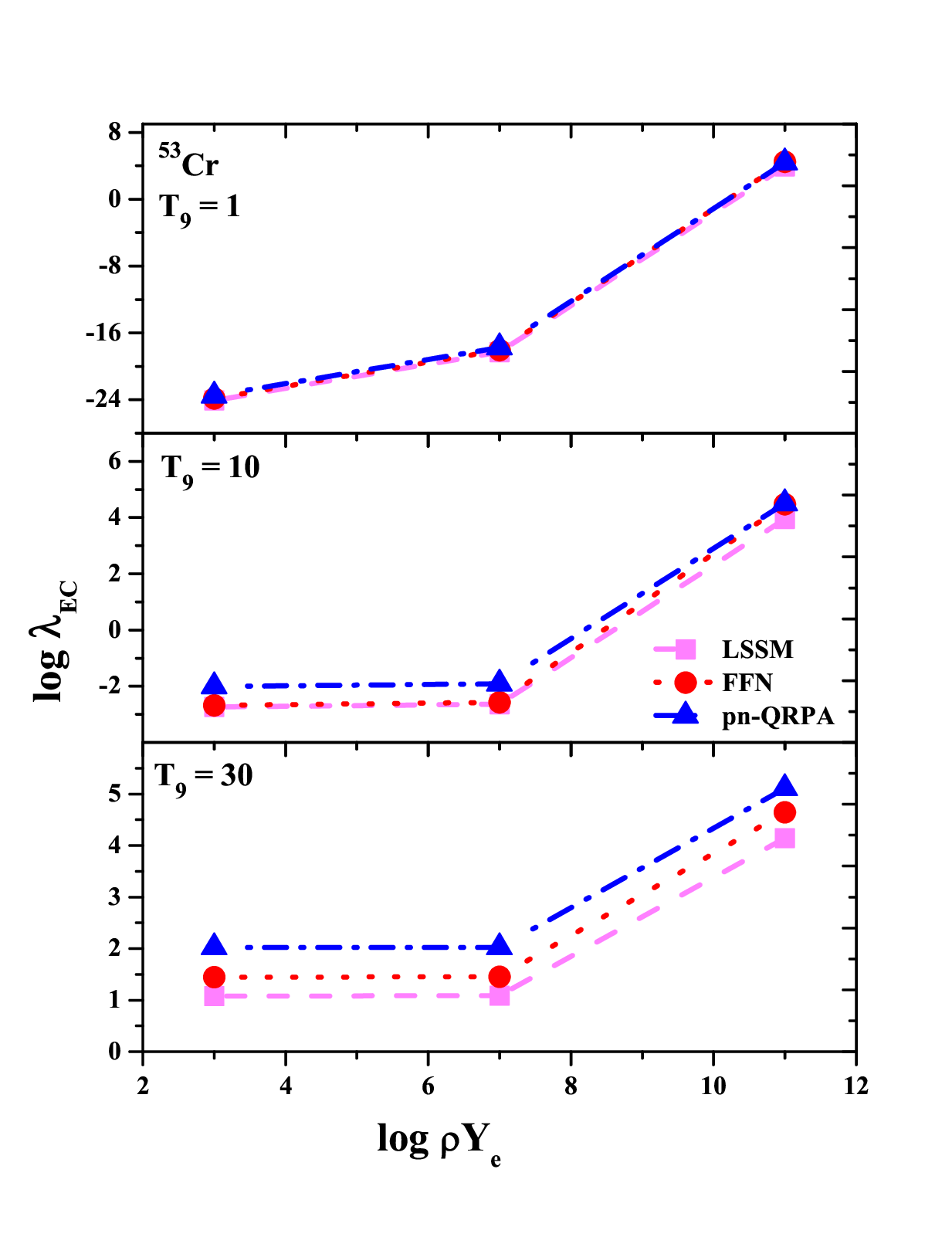} &
    \includegraphics[scale=0.33]{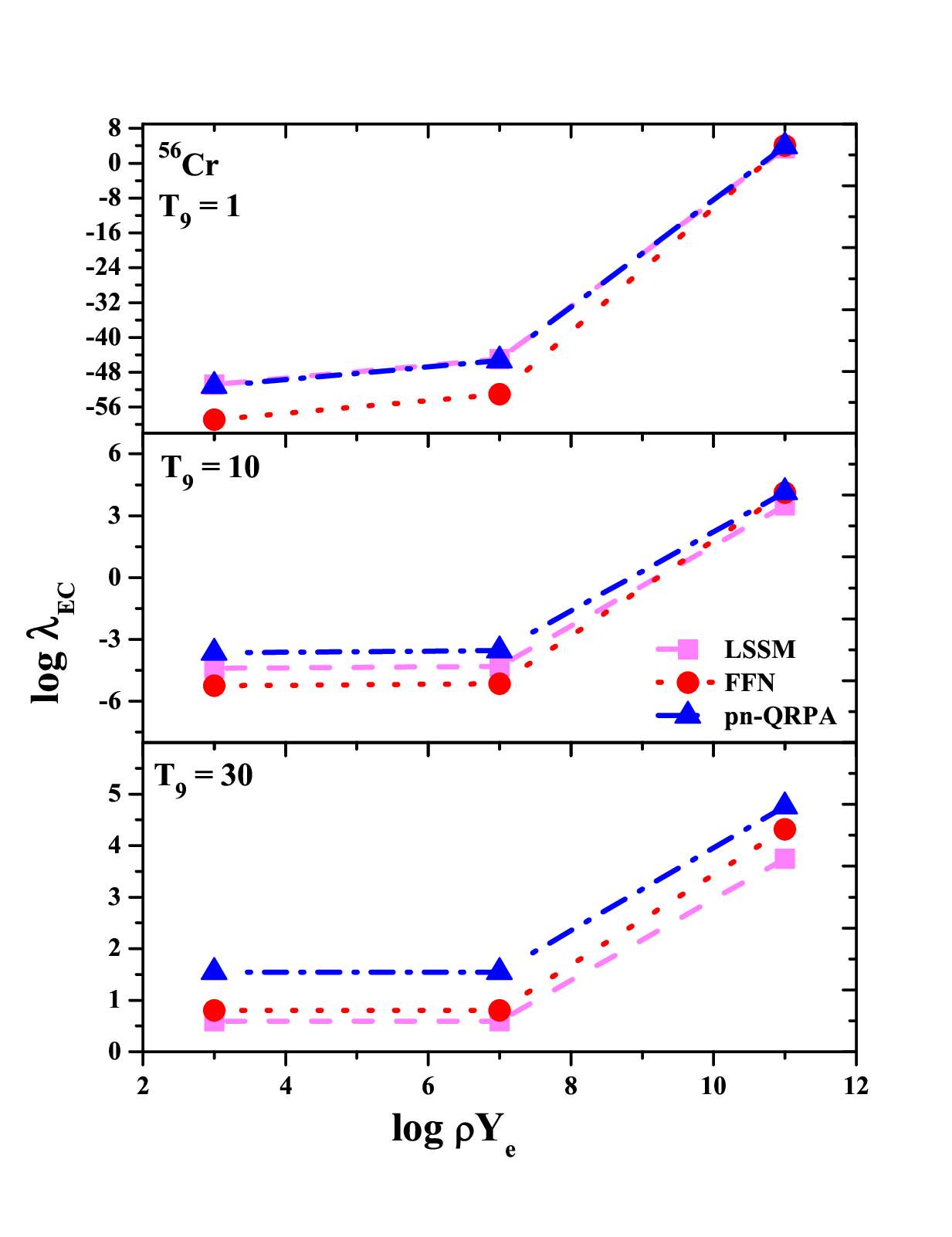}\\
    \includegraphics[scale=0.33]{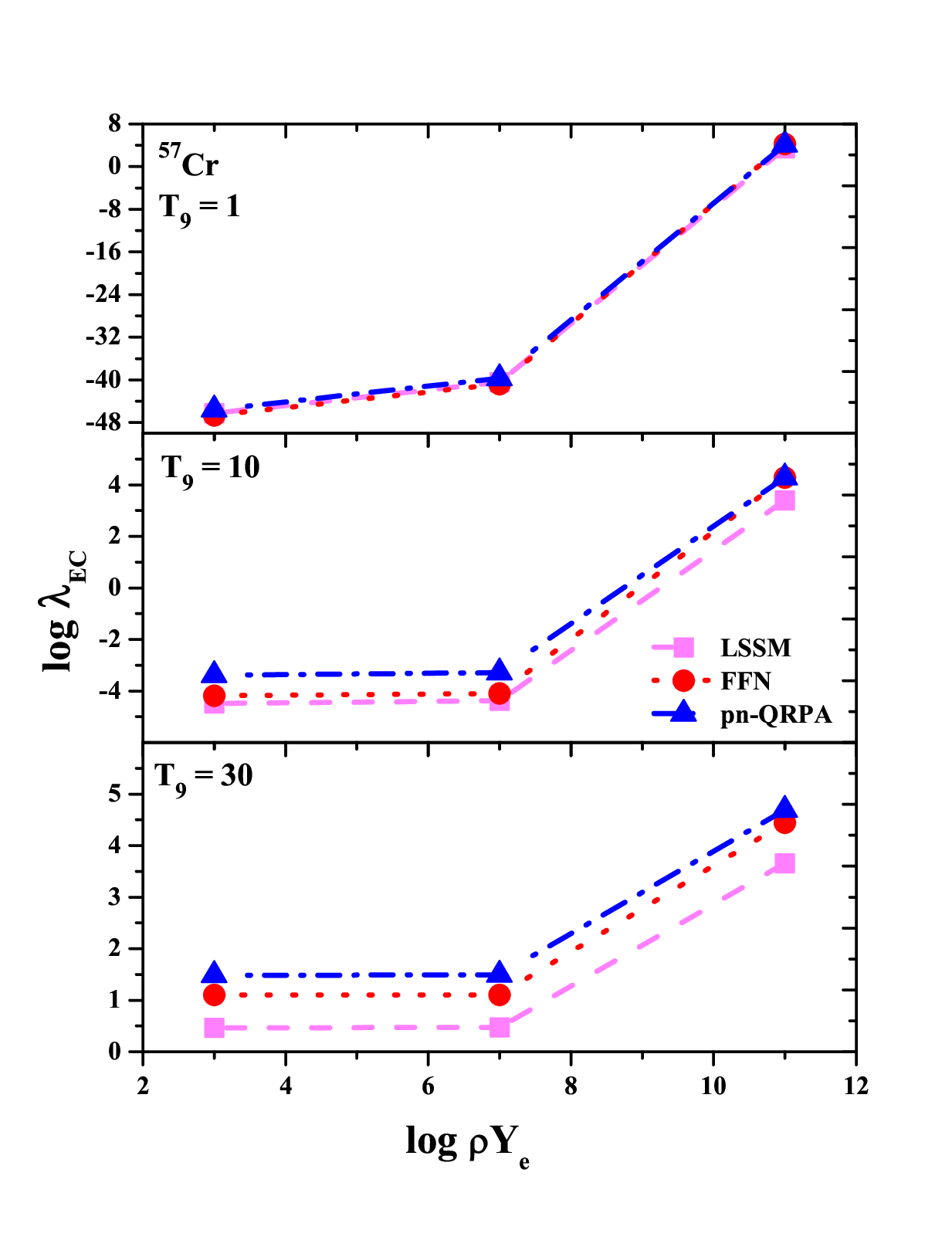} &
    \includegraphics[scale=0.33]{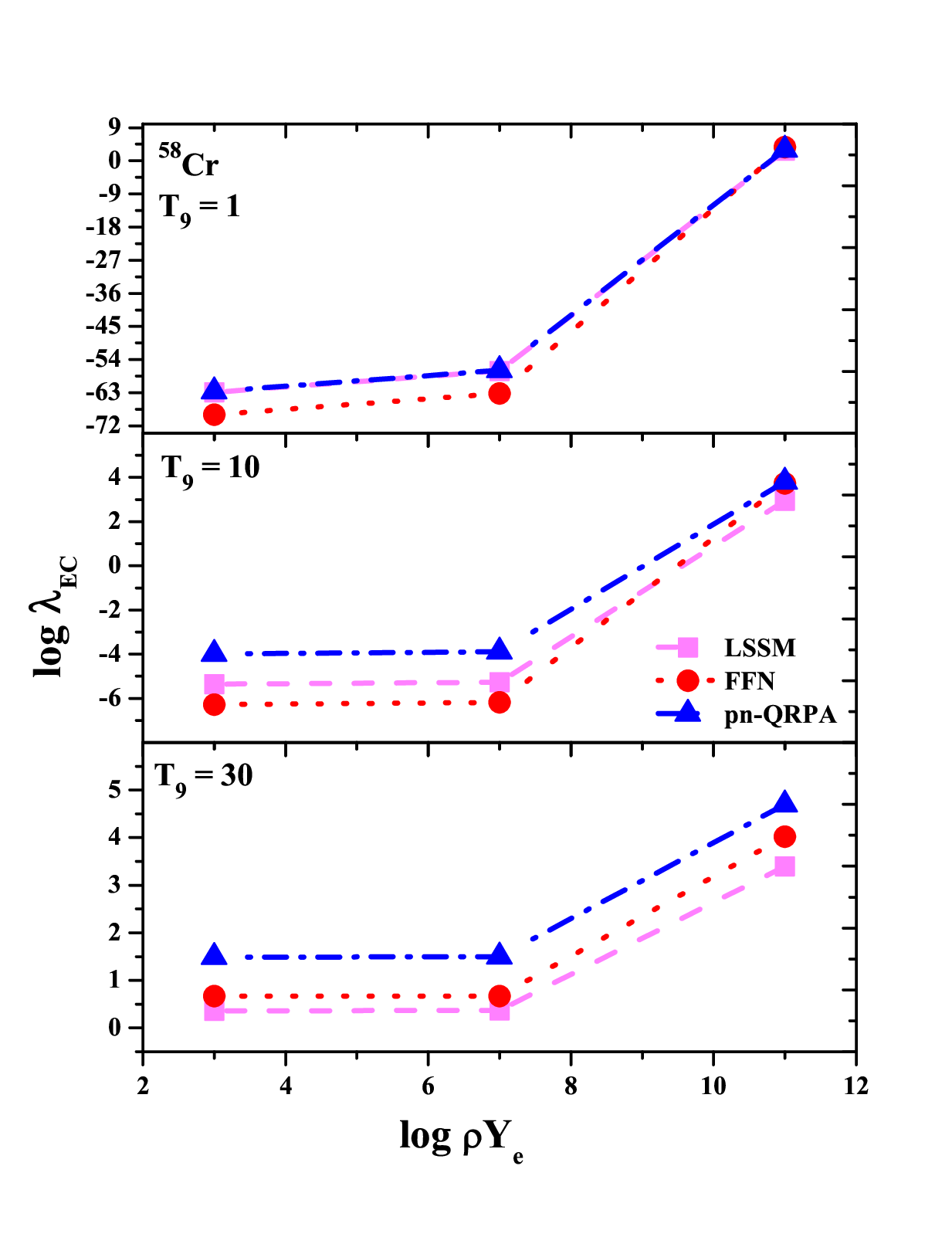}\\
 \end{tabular}
\caption{Same as Fig. 4, but for $^{53, 56-58}$Cr
isotopes.}\label{fig5}
\end{center}
\end{figure}

\clearpage
\begin{table}
\scriptsize \caption{Calculated EC rates in stellar region on
chromium ( $^{42-65}$Cr) isotopes at various densities and
temperatures. The first column shows the stellar densities
($\rho$Y$_{e}$) (in units of g/cm$^{3}$), where $\rho$ is the baryon
density and Y$_{e}$ is the ratio of the electron number to the
baryon number. T$_{9}$ are given in units of 10$^{9}$ K. The
calculated EC rates are tabulated in log to base 10 scale and given
in units of s$^{-1}$.} \label{Table 1}
\begin{center}
\scalebox{0.78}{
\begin{tabular} {cccccccccccccc}

    $\log\rho Y_{e}$ & T$_{9}$ & $^{42}$Cr & $^{43}$Cr & $^{44}$Cr & $^{45}$Cr & $^{46}$Cr  & $^{47}$Cr  & $^{48}$Cr  & $^{49}$Cr  & $^{50}$Cr & $^{51}$Cr & $^{52}$Cr & $^{53}$Cr \\
\hline
    3      & 1      & -3.239 & -3.396 & -3.428 & -3.383 & -3.935 & -4.294 & -5.917 & -5.674 & -13.029 & -5.104 & -26.535 & -23.539 \\
    3      & 3      & -0.941 & -1.057 & -1.119 & -1.090 & -1.609 & -1.848 & -3.314 & -3.062 & -5.872 & -2.401 & -10.061 & -8.695 \\
    3      & 10     & 0.990  & 0.946  & 0.860  & 0.848  & 0.489  & 0.252  & -0.571 & -0.501 & -1.341 & 0.141  & -2.246 & -2.010 \\
    3      & 30     & 3.054  & 3.343  & 3.158  & 3.215  & 2.803  & 2.466  & 2.274  & 2.373  & 1.999  & 2.934  & 1.802  & 2.021 \\
    7      & 1      & 0.463  & 0.305  & 0.283  & 0.327  & -0.209 & -0.576 & -1.954 & -1.818 & -7.209 & -0.924 & -20.724 & -17.736 \\
    7      & 3      & 0.474  & 0.358  & 0.301  & 0.331  & -0.180 & -0.425 & -1.789 & -1.572 & -4.159 & -0.833 & -8.358 & -6.984 \\
    7      & 10     & 1.082  & 1.038  & 0.953  & 0.940  & 0.582  & 0.344  & -0.475 & -0.405 & -1.244 & 0.238  & -2.148 & -1.911 \\
    7      & 30     & 3.058  & 3.347  & 3.161  & 2.806  & 3.218  & 2.470  & 2.278  & 2.377  & 2.003  & 2.938  & 1.806  & 2.024 \\
    11     & 1      & 5.360  & 5.263  & 5.338  & 5.375  & 5.098  & 4.860  & 4.811  & 4.688  & 4.611  & 5.625  & 4.510  & 4.394 \\
    11     & 3      & 5.361  & 5.273  & 5.338  & 5.389  & 5.099  & 4.910  & 4.812  & 4.726  & 4.611  & 5.627  & 4.511  & 4.396 \\
    11     & 10     & 5.385  & 5.338  & 5.364  & 5.395  & 5.153  & 5.081  & 4.832  & 4.774  & 4.631  & 5.630  & 4.539  & 4.498 \\
    11     & 30     & 5.633  & 5.905  & 5.775  & 5.478  & 5.862  & 5.202  & 5.120  & 5.254  & 4.949  & 5.762  & 4.889  & 5.112 \\

\hline

    $\log\rho Y_{e}$ & T$_{9}$    & $^{54}$Cr  & $^{55}$Cr  & $^{56}$Cr  & $^{57}$Cr  & $^{58}$Cr  & $^{59}$Cr  & $^{60}$Cr  & $^{61}$Cr  & $^{62}$Cr & $^{63}$Cr & $^{64}$Cr & $^{65}$Cr \\
\hline
    3      & 1      & -40.534 & -34.933 & -51.155 & -45.571 & -62.707 & -55.426 & -73.862 & -68.342 & -86.371 & -75.944 & -95.139 & -81.490 \\
    3      & 3      & -13.788 & -11.869 & -17.515 & -15.417 & -20.408 & -18.451 & -23.950 & -23.113 & -28.681 & -26.338 & -30.957 & -27.670 \\
    3      & 10     & -2.952 & -2.403 & -3.647 & -3.386 & -3.992 & -3.894 & -4.995 & -5.746 & -7.168 & -7.703 & -7.066 & -7.735 \\
    3      & 30     & 1.634  & 1.895  & 1.540  & 1.485  & 1.491  & 1.392  & 1.263  & 0.449  & -0.212 & -0.712 & 0.593  & -0.523 \\
    7      & 1      & -34.714 & -29.113 & -45.335 & -39.751 & -56.887 & -49.606 & -68.042 & -62.522 & -80.552 & -70.124 & -89.319 & -75.670 \\
    7      & 3      & -12.073 & -10.153 & -15.799 & -13.702 & -18.693 & -16.736 & -22.235 & -21.398 & -26.965 & -24.622 & -29.241 & -25.955 \\
    7      & 10     & -2.854 & -2.304 & -3.548 & -3.288 & -3.893 & -3.795 & -4.897 & -5.647 & -7.069 & -7.604 & -6.967 & -7.636 \\
    7      & 30     & 1.638  & 1.898  & 1.543  & 1.489  & 1.495  & 1.395  & 1.266  & 0.453  & -0.209 & -0.708 & 0.596  & -0.520 \\
    11     & 1      & 4.423  & 4.394  & 3.829  & 4.080  & 2.985  & 3.449  & 2.578  & 3.000  & 2.403  & 2.210  & 1.766  & 2.516 \\
    11     & 3      & 4.424  & 4.483  & 3.831  & 4.180  & 2.988  & 3.396  & 2.583  & 3.297  & 2.413  & 2.313  & 1.787  & 2.512 \\
    11     & 10     & 4.458  & 4.667  & 4.137  & 4.283  & 3.791  & 3.715  & 3.161  & 3.692  & 3.013  & 2.843  & 3.054  & 2.655 \\
    11     & 30     & 4.819  & 5.062  & 4.753  & 4.691  & 4.696  & 4.619  & 4.585  & 3.982  & 3.422  & 3.009  & 4.212  & 3.215 \\

\end{tabular}}
\end{center}
\end{table}

\begin{table}[h]
  \centering
\scriptsize  \caption{The pn-QRPA calculated total B(GT) strengths,
centroids and widths of Cr isotopes in electron capture
direction.}\label{Table 2}
    \begin{tabular}{c|c|c|c}

    Nuclei & $\sum$ B(GT$_{+}$) & $\bar{E}_{+}$ (MeV)  & Width$_{+}$ (MeV) \\
    \hline
    $^{42}$Cr & 7.08  & 6.80   & 3.25 \\
    $^{43}$Cr & 5.95  & 9.74  & 3.10 \\
    $^{44}$Cr & 5.33  & 7.96  & 5.55 \\
    $^{45}$Cr & 4.24  & 7.95  & 3.09 \\
    $^{46}$Cr & 4.31  & 4.76  & 3.31 \\
    $^{47}$Cr & 3.24  & 8.81  & 3.26 \\
    $^{48}$Cr & 3.33  & 4.19  & 2.56 \\
    $^{49}$Cr & 2.23  & 7.90   & 2.04 \\
    $^{50}$Cr & 2.49  & 4.03  & 2.41 \\
    $^{51}$Cr & 1.87  & 7.96  & 2.41 \\
    $^{52}$Cr & 2.21  & 3.23  & 2.01 \\
    $^{53}$Cr & 0.51  & 6.21  & 2.71 \\
    $^{54}$Cr & 1.95  & 2.11  & 3.68 \\
    $^{55}$Cr & 0.39  & 4.06  & 3.47 \\
    $^{56}$Cr & 1.31  & 1.77  & 2.14 \\
    $^{57}$Cr & 0.25  & 5.21  & 2.84 \\
    $^{58}$Cr & 0.82  & 1.57  & 2.49 \\
    $^{59}$Cr & 0.24  & 1.26  & 2.24 \\
    $^{60}$Cr & 0.39  & 3.03  & 4.99 \\
    $^{61}$Cr & 0.21  & 3.79  & 3.41 \\
    $^{62}$Cr & 0.23  & 3.22  & 5.51 \\
    $^{63}$Cr & 0.17  & 1.83  & 2.69 \\
    $^{64}$Cr & 0.16  & 2.63  & 5.06 \\
    $^{65}$Cr & 0.12  & 2.87  & 3.31 \\

    \end{tabular}
    \end{table}

    \begin{table}
\centering \scriptsize \caption{The pn-QRPA calculated centroids for
nickel  isotopes compared with other theoretical calculations and
measurements.  Centroid energies are given in units of MeV. For
references see text.}\label{Table 3}
    \begin{tabular}{c|c|c|c|c|c}
     Nucleus & $\bar{E}_{+}$ [pn-QRPA] & $\bar{E}_{+}$[PF] & $\bar{E}_{+}$[FFN] & $\bar{E}_{+}$[LSSM]&$\bar{E}_{+}$[exp] \\
    \hline
$^{58}$Ni & 3.57  & 3.65  & 3.76  & 3.75  & 3.60$\pm$0.20 \\
$^{60}$Ni & 3.09  & 2.70  & 2.00  & 2.88  & 2.40$\pm$0.30 \\
$^{61}$Ni & 4.93  & 4.70  & -  & 4.70  & - \\
$^{62}$Ni & 2.13  & 1.80  & 2.00  & 1.78  & 1.30$\pm$0.30 \\
$^{64}$Ni & 0.80  & 1.80  & 2.00  & 0.50  & 0.80$\pm$0.30 \\

    \end{tabular}
    \end{table}

\begin{table}
\centering \scriptsize \caption{Ratio of calculated electron capture
(EC) rates to $\beta^{+}$-decay for different selected densities and
temperatures. The second column shows the stellar densities
($\rho$Y$_{e}$) (in units of g/cm$^{3}$). T$_{9}$ are given in units
of 10$^{9}$ K.}\label{Table 4}
    \begin{tabular}{c|c|c|c|c|c}

           Nucleus & $\rho$$\it Y_{e}$  & \multicolumn{4}{c}{$R(EC/\beta^{+})$}\\
\cline{3-6} & &T$_{9}$=01 & T$_{9}$=05 & T$_{9}$=10 & T$_{9}$=30 \\
\hline

           & $10^{7}$ & 6.6E-02 & 7.5E-02 & 2.2E-01 & 4.4E+00 \\
    $^{42}$Cr & $10^{9}$ & 1.0E+01 & 1.1E+01 & 1.0E+01 & 8.6E+00 \\
           & $10^{11}$ & 5.2E+03 & 5.2E+03 & 4.4E+03 & 1.5E+03 \\
           \hline
           & $10^{7}$ & 4.2E-02 & 4.0E-02 & 8.9E-02 & 2.5E+00 \\
    $^{43}$Cr & $10^{9}$ & 6.8E+00 & 5.7E+00 & 3.9E+00 & 4.8E+00 \\
           & $10^{11}$ & 3.8E+03 & 2.9E+03 & 1.8E+03 & 8.4E+02 \\
           \hline
           & $10^{7}$ & 1.2E-01 & 1.4E-01 & 4.3E-01 & 1.2E+01 \\
    $^{44}$Cr & $10^{9}$ & 2.1E+01 & 2.2E+01 & 2.1E+01 & 2.3E+01 \\
           & $10^{11}$ & 1.3E+04 & 1.3E+04 & 1.1E+04 & 4.4E+03 \\
           \hline
           & $10^{7}$ & 1.8E-01 & 2.1E-01 & 7.3E-01 & 2.3E+01 \\
    $^{45}$Cr & $10^{9}$ & 3.4E+01 & 3.4E+01 & 3.5E+01 & 4.3E+01 \\
           & $10^{11}$ & 2.2E+04 & 2.1E+04 & 2.0E+04 & 8.7E+03 \\
           \hline
           & $10^{7}$ & 2.3E-01 & 2.9E-01 & 9.1E-01 & 4.4E+01 \\
    $^{46}$Cr & $10^{9}$ & 5.4E+01 & 5.6E+01 & 4.7E+01 & 8.2E+01 \\
           & $10^{11}$ & 4.7E+04 & 4.7E+04 & 3.3E+04 & 1.7E+04 \\
           \hline
           & $10^{7}$ & 2.5E-01 & 3.0E-01 & 1.2E+00 & 9.1E+01 \\
    $^{47}$Cr & $10^{9}$ & 5.4E+01 & 5.3E+01 & 5.9E+01 & 1.7E+02 \\
           & $10^{11}$ & 6.9E+04 & 5.3E+04 & 5.0E+04 & 3.9E+04 \\
           \hline
           & $10^{7}$ & 1.5E+03 & 4.6E+02 & 2.0E+01 & 3.5E+02 \\
    $^{48}$Cr & $10^{9}$ & 2.1E+06 & 2.8E+05 & 1.7E+03 & 6.5E+02 \\
           & $10^{11}$ & 8.6E+09 & 1.0E+09 & 3.9E+06 & 1.9E+05 \\
           \hline
           & $10^{7}$ & 2.9E+01 & 3.3E+01 & 6.1E+01 & 2.0E+03 \\
    $^{49}$Cr & $10^{9}$ & 2.6E+04 & 1.4E+04 & 4.2E+03 & 3.6E+03 \\
           & $10^{11}$ & 9.2E+07 & 3.8E+07 & 8.5E+06 & 1.1E+06 \\
           \hline
           & $10^{7}$ & 2.9E+14 & 8.8E+02 & 7.9E+01 & 1.8E+03 \\
    $^{50}$Cr & $10^{9}$ & 9.6E+21 & 2.8E+06 & 8.3E+03 & 3.3E+03 \\
           & $10^{11}$ & 1.9E+26 & 4.1E+10 & 5.4E+07 & 1.2E+06 \\
           \hline
           & $10^{7}$ & 6.5E+07 & 9.0E+04 & 6.7E+03 & 2.3E+04 \\
    $^{51}$Cr & $10^{9}$ & 1.1E+11 & 5.0E+07 & 4.9E+05 & 4.2E+04 \\
           & $10^{11}$ & 2.3E+14 & 9.4E+10 & 1.5E+09 & 1.2E+07 \\
           \hline
           & $10^{7}$ & 5.2E+11 & 5.1E+03 & 6.5E+02 & 8.8E+03 \\
    $^{52}$Cr & $10^{9}$ & 2.2E+30 & 7.7E+07 & 8.1E+04 & 1.6E+04 \\
           & $10^{11}$ & 8.9E+36 & 4.2E+13 & 2.9E+09 & 8.1E+06 \\
           \hline
           & $10^{7}$ & 1.2E+23 & 1.2E+06 & 3.7E+04 & 2.1E+06 \\
    $^{53}$Cr & $10^{9}$ & 5.2E+38 & 4.6E+09 & 4.1E+06 & 3.6E+06 \\
           & $10^{11}$ & 1.6E+45 & 1.0E+15 & 8.4E+10 & 1.7E+09 \\
           \hline
           & $10^{7}$ & 9.5E+14 & 1.2E+06 & 3.3E+05 & 3.2E+07 \\
    $^{54}$Cr & $10^{9}$ & 1.0E+35 & 2.7E+10 & 4.7E+07 & 5.3E+07 \\
           & $10^{11}$ & 1.3E+54 & 5.2E+17 & 5.5E+12 & 2.9E+10 \\

\end{tabular}
\end{table}

\end{document}